\documentclass[]{emulateapj}

\usepackage{natbib}

\usepackage{epsfig}
\usepackage{color}
\usepackage{graphicx}

\newcommand{\rxte}{{\it RXTE\/}}

\newcommand{\maxi}{{\it MAXI\/}}
\newcommand{\rosat}{{\it ROSAT\/}}

\newcommand{\xmm}{{\it XMM-Newton}}

\newcommand{\swift}{{\it Swift\/}}

\newcommand{\nustar}{\textit{NuSTAR\/}}
\newcommand{\nicer}{\textit{NICER\/}}

\newcommand\etal{{et al.}}

\def\xte{XTE~J1810$-$197}

\def\farcm{\hbox{$.\mkern-4mu^\prime$}}
\def\farcs{\hbox{$.\!\!^{\prime\prime}$}}

\begin{document}

\shorttitle{2018 Outburst of \xte}

\shortauthors{Gotthelf \etal}

%\title{The 2018 X-ray Outburst of Magnetar \xte}
\title{The 2018 X-ray and Radio Outburst of Magnetar \xte}
%\title{The 2018 X-ray and Radio Outburst of Magnetar \xte\ and its Radio Phase Alignment}

\author{E.~V.~Gotthelf\altaffilmark{1}, J.~P.~Halpern\altaffilmark{1}, J.~A.~J.~Alford\altaffilmark{1},
T.~Mihara\altaffilmark{2}, H.~Negoro\altaffilmark{3}, N.~Kawai\altaffilmark{4},
S.~Dai\altaffilmark{5}, M.~E.~Lower\altaffilmark{6},  S.~Johnston\altaffilmark{5}, M.~Bailes\altaffilmark{6}, S.~Os{\l}owski\altaffilmark{6}, F.~Camilo\altaffilmark{7}, H.~Miyasaka\altaffilmark{8}, K.~K.~Madsen\altaffilmark{8}}
\altaffiltext{1}{Columbia Astrophysics Laboratory, Columbia University, 550 West 120th Street, New York, NY 10027, USA}
\altaffiltext{2}{High Energy Astrophysics Laboratory, RIKEN, 2-1 Hirosawa, Wako, Saitama 351-0198, Japan}
\altaffiltext{3}{Department of Physics, Nihon University, 1-8 Kanda-Surugadai, Chiyoda-ku, Tokyo 101-8308, Japan}
\altaffiltext{4}{Department of Physics, Tokyo Institute of Technology, 2-12-1 Ookayama, Meguro-ku, Tokyo 152-8551, Japan}
\altaffiltext{5}{CSIRO Astronomy and Space Science, Australia Telescope National Facility, PO Box 76, Epping, NSW 1710, Australia}
\altaffiltext{6}{Centre for Astrophysics and Supercomputing, Swinburne University of Technology, VIC 3122, Australia}
\altaffiltext{7}{South African Radio Astronomy Observatory, Observatory 7925, South Africa}
\altaffiltext{8}{Cahill Center for Astronomy and Astrophysics, California Institute of Technology, Pasadena, CA 91125, USA}
%\author{F. A.~Harrison, K.~Forster, B.~Grefenstette, K. C.~Madsen, H.~Miyasaka}
%\affil{Cahill Center for Astronomy and Astrophysics, California Institute of Technology, Pasadena, CA 91125, USA}

\begin{abstract}
  We present the earliest X-ray observations of the 2018 outburst of
  \xte, the first outburst since its 2003 discovery as the
  prototypical transient and radio-emitting anomalous X-ray pulsar
  (AXP).  The \textit{Monitor of All-sky X-ray Image} (\maxi\/)
  detected \xte\ immediately after a November 20--26 visibility gap,
  contemporaneous with its reactivation as a radio pulsar, first
  observed on December 8.  On December 13 the \textit{Nuclear
    Spectroscopic Telescope Array} (\nustar) detected X-ray emission
  up to at least 30~keV, with a spectrum well-characterized by a
  blackbody plus power-law model with temperature $kT =
  0.74\pm0.02$~keV and photon index $\Gamma = 4.4\pm0.2$ or by a
  two-blackbody model with $kT = 0.59\pm0.04$~keV and $kT =
  1.0\pm0.1$~keV, both including an additional power-law component to
  account for emission above 10~keV, with $\Gamma_h = -0.2\pm1.5$ and
  $\Gamma_h = 1.5\pm0.5$, respectively. The latter index is consistent with
  hard X-ray flux reported for the non-transient magnetars.  In the
  2$-$10~keV bandpass, the absorbed flux is $2\times10^{-10}$
  erg~s$^{-1}$~cm$^{-2}$, a factor of 2 greater than the maximum flux
  extrapolated for the 2003 outburst.  The peak of the sinusoidal
  X-ray pulse lags the radio pulse by $\approx0.13$ cycles, consistent
  with their phase relationship during the 2003 outburst.  This
  suggests a stable geometry in which radio emission originates on
  magnetic field lines containing currents that heat a spot on the
  neutron star surface.  However, a measured energy-dependent phase
  shift of the pulsed X-rays suggests that all X-ray emitting regions
  are not precisely co-aligned.
%This phase spread may be transitory, as we found
%no such effect at later times in the 2003 outburst.
% It is now possible to study the origin
%  and decay of magnetar emission in exquisite detail from the
%  beginning of the outburst in a uniquely accessible source.
\end{abstract}

%These results are generally comparable to those found during the 2003 outburst.  

%An equivalent fit is  obtained for a two blackbody model with $kT = 1.2$~keV and $kT =  0.62$~keV.

\keywords{pulsars: general --- pulsars: individual (\xte) --- stars: neutron --- X-rays: stars}

\section{INTRODUCTION}

Magnetars are neutron star (NS) pulsars whose X-ray luminosity can
greatly exceed their spin-down power.  Unlike for the canonical
rotation-powered radio pulsars, the luminosity of magnetars is thought
to be supplied by the decay of their large magnetic fields, typically
$\approx10^{14-15} $ G.  The dipole field components result in rapid
spin-down and long rotation periods in the 0.3--12~s range.  The
discovery of \xte\ \citep{ibr04,got04} marked a turning point in the
study of magnetars.  Until then, the known magnetars comprised four
transient soft gamma-ray repeaters (SGRs), and five persistent
anomalous X-ray pulsars (AXPs) detected by {\it UHURU\/}, {\it
  Einstein}, or \rosat.  The two classes shared similar magnetic field
strengths and spin periods but had different long-term histories.  The
SGRs had rare, violent outbursts, while the AXPs were fairly steady
emitters. See recent reviews by \cite{kas17,esp18,cot18}.

\xte\ was the first recognized transient AXP, detected in outburst in 2003 with a
period of 5.54~s at a flux level $\sim$140 times higher than its
quiescent state as a previously anonymous \rosat\ source.  It decayed
roughly as a $\tau\approx280$ day exponential.  A VLA survey in 2004
serendipitously detected a point source at the position of \xte\
\citep{hal05b}; subsequently pulsed radio emission was searched for and
detected for the first time in a magnetar \citep{cam06}.  Until then
it had been theorized that high magnetic fields inherently suppressed
radio pulsations.  On the contrary, \xte\ was a bright transient radio
pulsar, but with a flatter spectrum than ordinary pulsars, such that
it was the brightest neutron star known at frequencies above 20~GHz.
Three more transient magnetars have been detected as radio pulsars
with similar properties \citep{cam07a,lev10,sha13,eat13}.

Meanwhile, short SGR-like bursts had also been discovered from AXPs
\citep{gav02,kas03}, reinforcing the connection between the two
magnetar classes.  Similar bursts were detected during the decay of
\xte\ \citep{woo05}.  Beginning with the launch of \swift/BAT,
transient magnetars have been discovered on a regular basis, each outburst
signaled by one or more short SGR-like bursts.  There are now a total
of 23 confirmed
magnetars\footnote{http://www.physics.mcgill.ca/$\sim$pulsar/magnetar/main.html}.
Interestingly, no new {\it persistent\/} ones have been discovered
since 2007, suggesting that most magnetars are transient.

\xte\ provides a crucial probe of NS surface physics because of its
relatively close distance \citep[3$-$4~kpc][]{min08,dur06} and
continuing quiescent emission, not detected in most other transient
magnetars. This allows the evolving spectrum and pulse profiles of
\xte\ to be modeled \citep{per08,alb10,ber11}, mapping the magnetar's
cooling and shrinking surface thermal hot spots, presumably powered by
currents along the untwisting magnetic field-line bundles \citep[``j-bundles'',][]{bel09,bel13}.

In the following sections we present the first X-ray observations of \xte\
during its 2018 outburst.  \maxi\ all-sky monitoring data
constrains the epoch of the outburst, and a timely \nustar\ observation
within $\approx3$ weeks of the onset characterizes its early spectrum 
in the 3$-$30~keV band.  A comparison of the X-ray pulse phase with a
contemporaneous radio pulse observation is also made.  We discuss our
results in the context of the previous outburst and future expectations.

\section{X-ray Observations}

The reports of intense radio emission \citep{lyn18,des18,low18} and
enhanced X-ray flux \citep{mih18} from \xte\ signaled a new outburst
from this magnetar that had occurred sometime between 2018 October~26
and December 8 (radio) and between 2018 November 20--26 (X-ray; this
work).  Based on the \citet{lyn18} discovery, we initiated a \nustar\
Director's Discretionary Time observation of the magnetar.
Preliminary \nustar\ results were reported in \citet{got18} using a
subset ($\sim$50\%) of the data.  In the current work, we analyze the
complete \nustar\ observation of \xte, along with the \maxi\ all-sky
monitoring light curve, to determine its early outburst spectral and
temporal properties.

In the following study, all spectra for \xte\ are fitted using XSPEC
v12.10.0c software (Arnaud 1996) with the column density characterized
by the default {\tt TBabs} absorption model. Spectral uncertainties
are computed for the 90\% confidence level for two interesting
parameters unless otherwise noted.
% selecting the {\tt wilm} Solar
%abundances \citep{wil00} and the {\tt vern} photoionization
%cross-section \cite{ver96}.  
For the timing analysis, photons arrival
times were converted to the solar system barycenter using the radio
coordinates and the JPL DE200 planetary ephemeris.

\subsection{\maxi\  Results}

\begin{figure}[t]
%\vspace{-0.3cm} 
\centerline{
\hfill
\psfig{figure=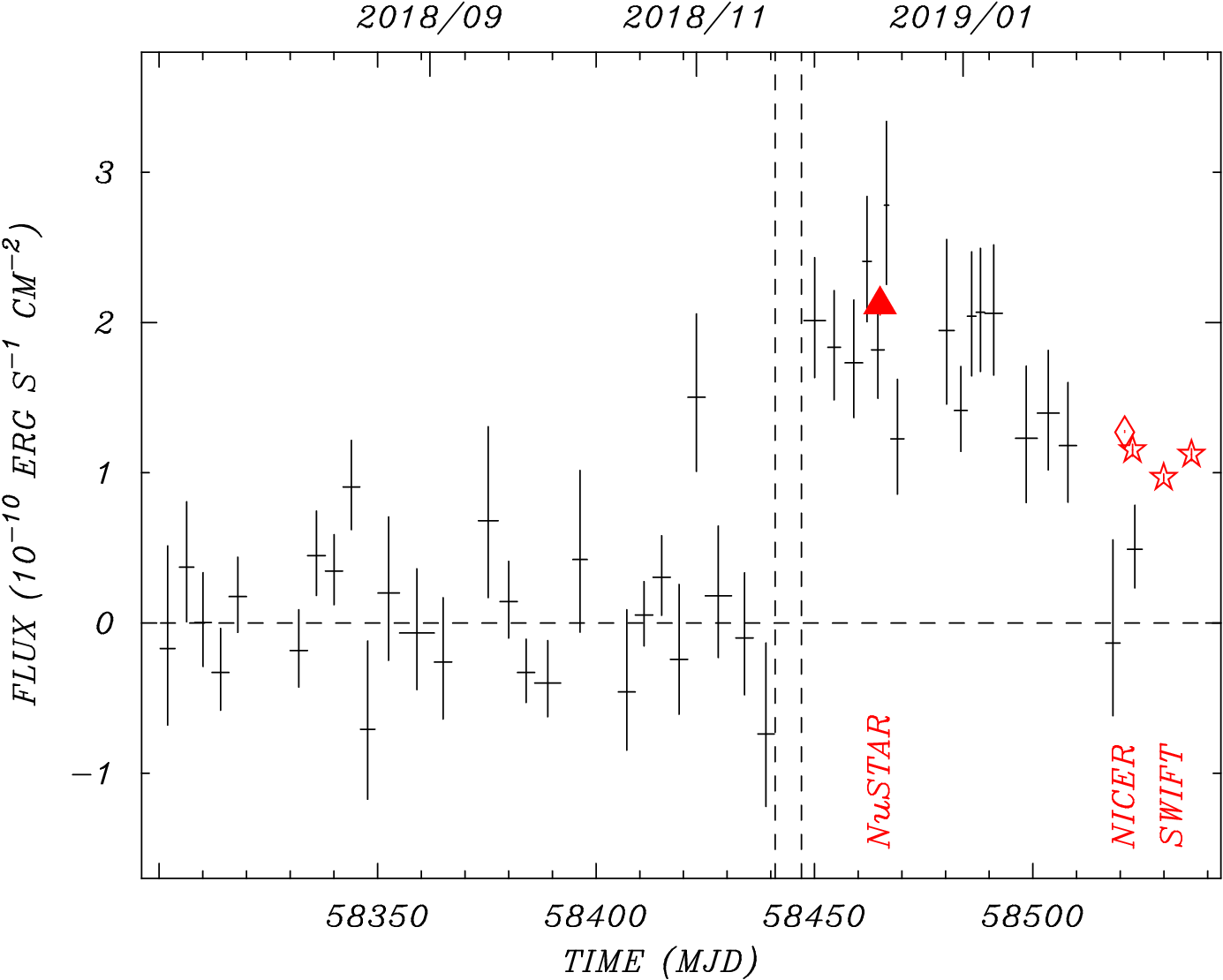,height=1.\linewidth,angle=270}
\hfill
}
%\vspace{-0.2cm} 
\caption{\footnotesize \maxi\ 2$-$10~keV light curve of \xte\ using a variable
  binning scheme (see text). The vertical lines bound the possible time of outburst.
% the last unambiguous non-detection and the first definitive detection.  
  The decrease in flux over time since outburst is evident and
  consistent with flux measurements obtained with \nustar\ (triangle;
  this work), \nicer\ \citep[diamond;][]{guv19} \&  \swift\ 
  (stars). The latter were obtained from spectra generated from reprocessed
  \swift\ archival data fitted with a blackbody model in the 1$-$5 keV range.}
\label{fig:maxilc} 
%\vspace{-0.3cm} 
\end{figure}

The \maxi\ observatory \citep{mat09} is attached to the International
Space Station (ISS) and scans the sky 16 times a day during its 92~min
orbit, using the Gas Slit Camera \citep[GSC,][]{mih11} to build up
images in the 2$-$30 keV band.  Although \xte\ is not a \maxi\
cataloged object \citep{hor18}, the region containing the magnetar was
observed with the GSC as part of the \maxi\ Nova-Alert System program
\citep{neg16}.  The region was observed before and after the reported
radio event for $\approx$ 40$-$140~s per scan, with data gaps due to
Earth-block, ISS structure obscuration, SAA passages, and periods of
high particle background. A light curve at the position of \xte\ was
extracted from stacked 1~day image scans using a $1\fdg6$ radius
aperture, excluding two slightly overlapping sources.  Starting on
November~26 (MJD 58448), we detect a significant increase in the 1~day
count rates, following a 6~day gap that lacked reliable image data.
During the interval MJD 58450 - 58498, the average 2$-$10~keV rate
increased to $0.0195\pm0.0013$~s$^{-1}$~cm$^{-2}$, compared to the
pre-outburst rate of $0.0012\pm 0.0062$~s$^{-1}$~cm$^{-2}$.
Figure~\ref{fig:maxilc} displays the 2$-$10~keV light curve in energy
flux units, rebinned to obtain at least a $5\sigma$ detection over a
maximum time span of 4 days.  MAXI count rates were converted to flux
units using the PIMMS\footnote{Portable, Interactive Multi-Mission
  Simulator;
  https://heasarc.gsfc.nasa.gov/docs/software/tools/pimms.html}
software, for a blackbody temperature of 0.7~keV, estimated from the
average 4$-$10 keV/2$-$4 keV hardness ratio during the MJD 58450 - 58498
interval.
We conclude that \xte\ became active in X-rays sometime between November 20--26.

%, compared to the fluxmeasurements obtained with \nustar\ (this work) and \nicer\ \citep{guv19}.  

%The GSC spectrum of the magnetar, extracted from data collected over 8
%days following the outburst is well-characterized by a simple
%blackbody model of $kT = 0.90\pm0.13$~keV, without interstellar
%absorption, with a 2$-$10~keV flux of $(1.6\pm0.2) \times
%10{^{-10}}$~erg~s$^{-1}$~cm$^{-2}$. Spectral fits using a power-law
%model yields a photon index $\Gamma = 2.7\pm0.4$ with a lower
%statistic significance. 

\subsection{\nustar\  Observation}

We obtained a 22~hour observation of \xte\ starting on UT 2018
December~13 at 03:10:21 UT. At this time, \nustar\ was the only X-ray
mission capable of imaging the source so close ($12^{\circ}$) to the
Sun. A single source is detected in the field of view, with a count
rate of 4.8 s${^{-1}}$ and flux up to at least 30~keV in the
3$-$79~keV band. Recovery of the expected 5.54~s NS spin period
identifies the source as \xte\ (see Section \ref{timing}).

\nustar\ consists of two co-aligned X-ray telescopes, with
corresponding focal plane detector modules FPMA and FPMB, each of
which is composed of a $2\times2$-node CdZnTe sensor array
\citep{Harrison2013}.  These are sensitive to X-rays in the 3$-$79~keV
band, with a characteristic spectral resolution of 400~eV FWHM at
10~keV. The multi-nested foil mirrors provide $18^{\prime\prime}$ FWHM
($58^{\prime\prime}$ HPD) imaging resolution over a $12\farcm2\times
12\farcm2$ field-of-view \citep{Harrison2013}.  The nominal timing
accuracy of \nustar\ is $\sim$2~ms rms, after correcting for drift of
the on-board clock, with the absolute timescale shown to be better
than $< $$3$~ms \citep{Mori14, Madsen15}.  This is more than
sufficient to resolve the X-ray signal from \xte\ and to compare it to
the radio pulse.

\begin{figure*}[t]
%\vspace{-0.3cm} 
\centerline{
\psfig{figure=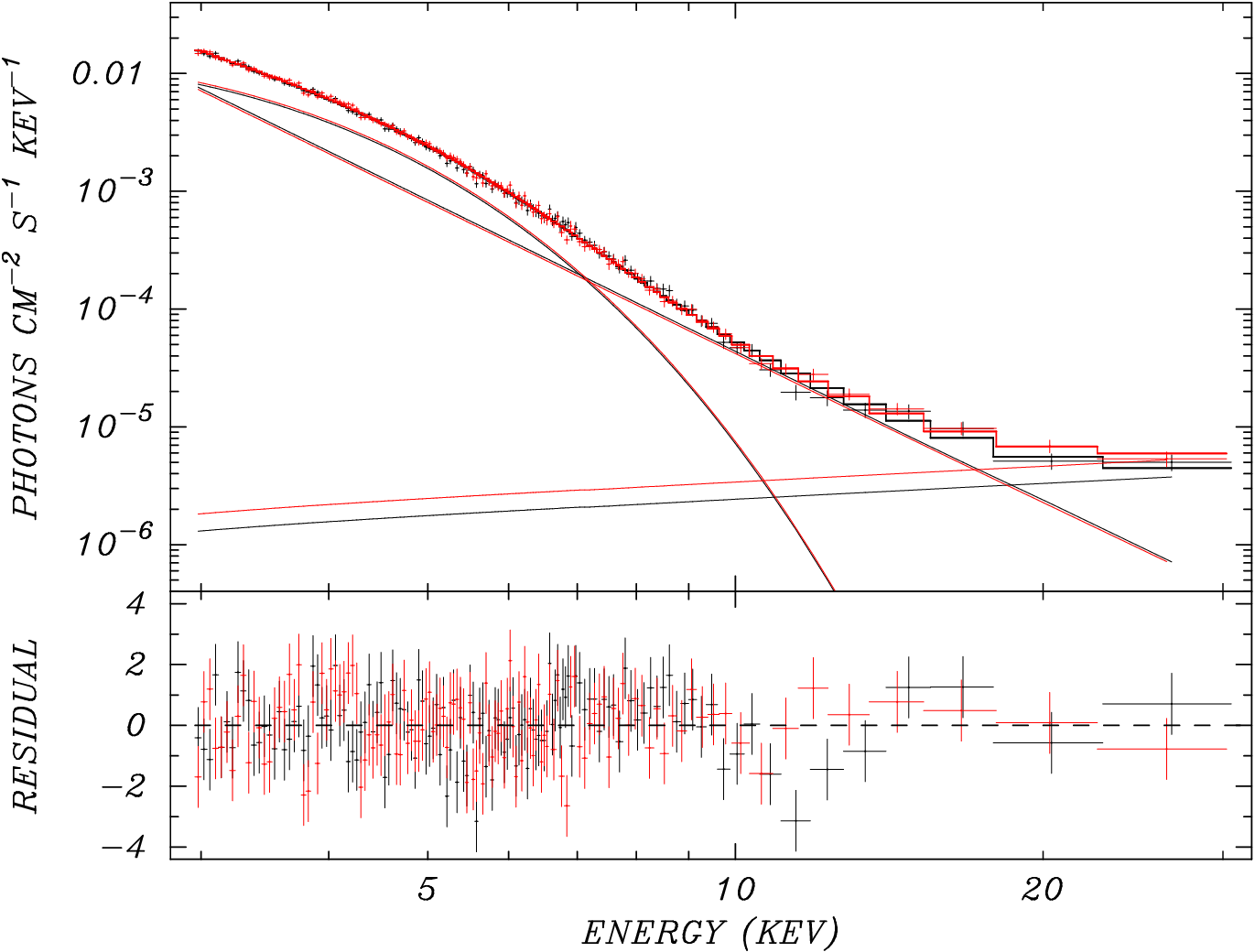,width=0.36\linewidth,angle=270}
\hfill
\psfig{figure=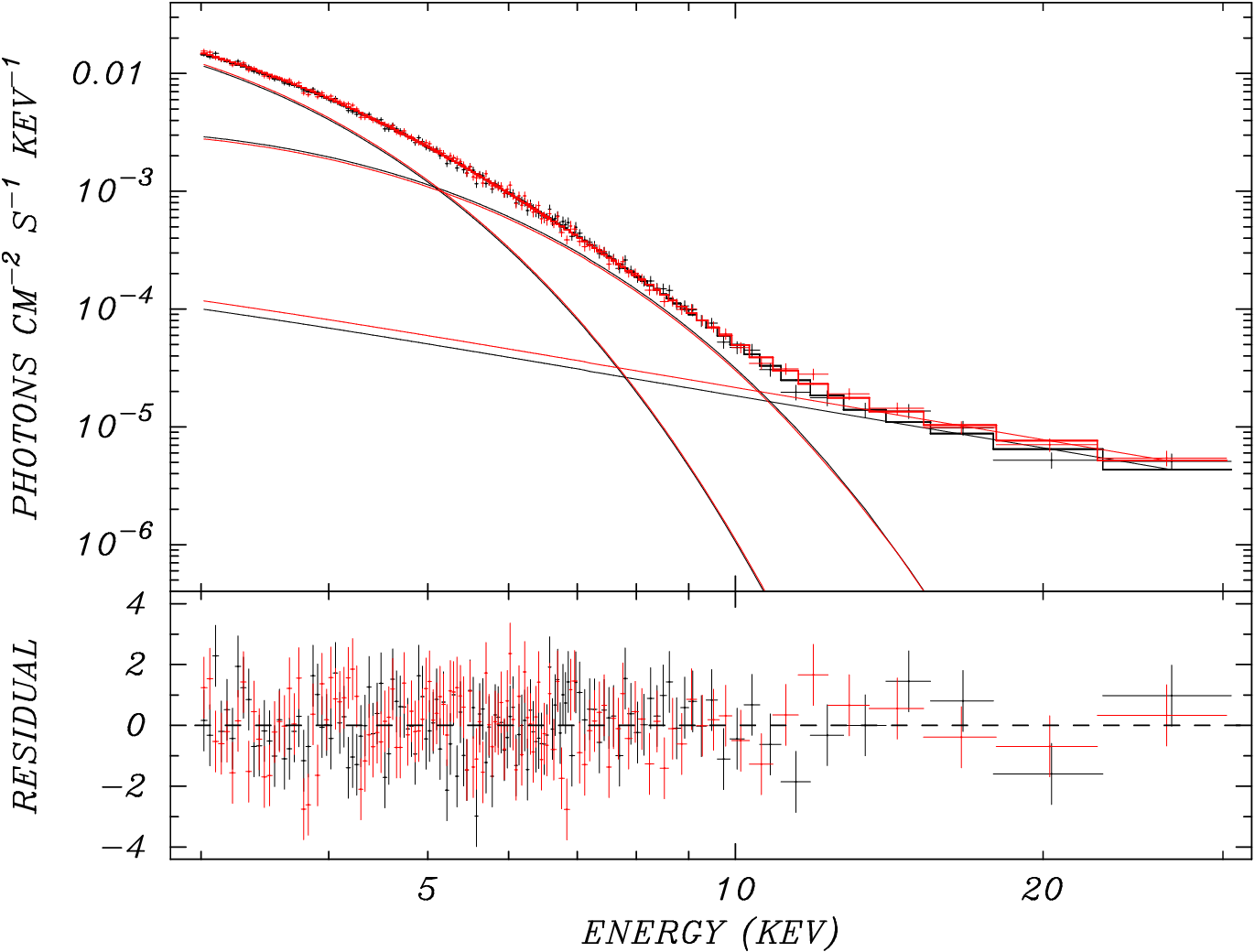,width=0.36\linewidth,angle=270}
\hfill
}
%\vspace{-0.2cm} 
\caption{ \footnotesize \nustar\ 3$-$30 ~keV spectrum of \xte\
  obtained during its December 2018 outburst fitted to the blackbody
  plus power-law model (Left) or two-blackbody model (Right) described in the text.  Both
  models include an additional power-law component to characterize the
  hard $>10$~keV flux.  The top panel shows the unfolded data
  (crosses) and fitted model (solid lines) collected by the two focal
  plane modules FPMA (black) and FPMB (red), along with their spectral
  components.  The lower panel shows the residual between the data and
  the model, in units of sigma.}
\label{fig:spectra} 
%\vspace{-0.3cm} 
\end{figure*}

\begin{figure}[t]
%\vspace{-0.3cm} 
\centerline{
\hfill
\psfig{figure=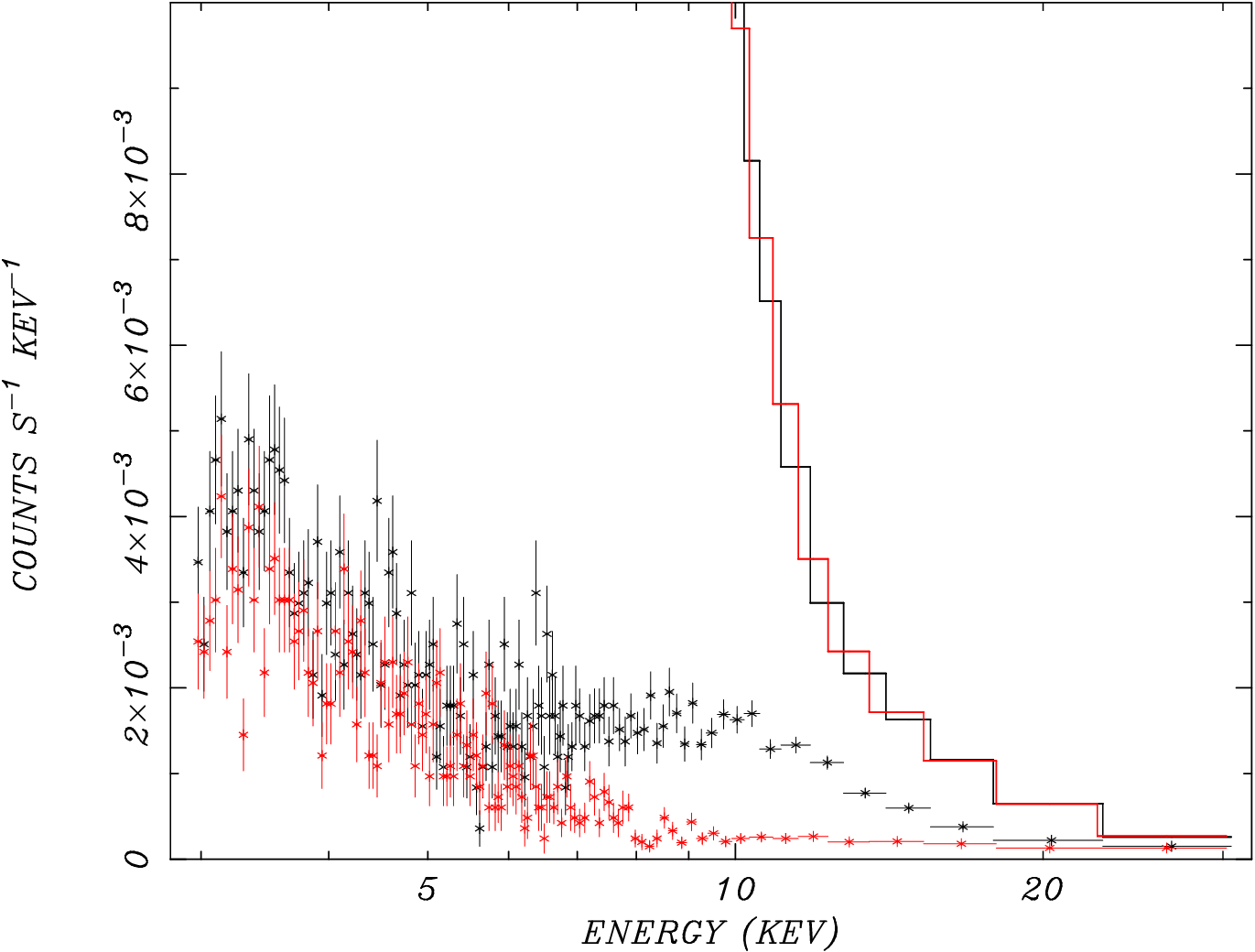,width=0.72\linewidth,angle=270}
\hfill
}
%\vspace{-0.2cm} 
\caption{ The  best fit blackbody
  plus power-law model (solid line) in the 3$-$30~kev band (see 
  Figure~\ref{fig:spectra}) compared with the background
  contribution (crosses), scaled linearly. The background rate is
  negligible below 10~keV and nowhere exceeds the source rate for all
  spectral channels in the full 3$-$30~keV energy range.}
\label{fig:specback} 
%\vspace{-0.3cm} 
\end{figure}

\begin{deluxetable}{lr}
\tablewidth{0pt}
%\tabletypesize{\scriptsize}
\tablecaption{\nustar\ 3-30~keV Spectral Fit Results\label{tab:spectra}}
\tablehead{
\colhead{Model}   & \colhead {Parameter}
}
\startdata
\cutinhead{Two-blackbody Model with Hard Power-law} \\
$N_{\rm H}$ ($10^{22}$ cm$^{-2}$)                      & $1.0$ (fixed)	\\
$kT_1$ (keV)					       & $0.59 \pm 0.04$	\\
$kT_2$ (keV)					       & $1.0 \pm 0.1$	\\
$\Gamma_h$					       & $1.5 \pm 0.5$	     \\
BB1 Flux (2$-$10~keV)\tablenotemark{a}                 & $1.4\times10^{-10}$\\
BB2 Flux (2$-$10~keV)\tablenotemark{a}                 & $5.3\times10^{-11}$  \\
PL Flux (2$-$10~keV)\tablenotemark{a}                  & $3.4\times10^{-12}$  \\
PL Flux (2$-$30~keV)\tablenotemark{a}                  & $8.1\times10^{-12}$  \\
Total Flux (2$-$10~keV)\tablenotemark{a}               & $1.9\times10^{-10}$\\
$L_{\rm BB1}$(bol) (erg s$^{-1}$)\tablenotemark{b}     & $4.3\times10^{35}$ 	\\
$L_{\rm BB2}$(bol) (erg s$^{-1}$)\tablenotemark{b}     & $1.0\times10^{35}$    \\
BB1 Area (cm$^2$)				       & $3.4\times10^{12}$	\\
BB2 Area (cm$^2$) 				       & $9.8\times10^{10}$	\\
$\chi^2_{\nu}$(dof)             		       & 1.04(250) \\    
\cutinhead{Power-law + Blackbody Model with Hard Power-law}\\
$N_{\rm H}$ ($10^{22}$ cm$^{-2}$)                      & $1.0$ (fixed)	     \\
$kT_{\rm BB}$ (keV)				       & $0.74 \pm 0.02$ 	     \\
$\Gamma_s$				               & $4.4 \pm 0.2$	     \\
$\Gamma_h$		     	                       & $-0.2 \pm 1.5$	     \\
BB Flux (2$-$10~keV)\tablenotemark{a}                  & $1.0\times10^{-10}$  \\
PL$_{\rm s}$ Flux (2$-$10~keV)\tablenotemark{a}        & $1.2\times10^{-10}$  \\
PL$_{\rm h}$ Flux (2$-$10~keV)\tablenotemark{a}        & $2.5\times10^{-13}$  \\
PL$_{\rm h}$ Flux (2$-$30~keV)\tablenotemark{a}        & $3.1\times10^{-12}$  \\
Total Flux (2$-$10~keV)\tablenotemark{a}               & $2.2\times10^{-10}$  \\
$L_{\rm BB}$(bol) (erg s$^{-1}$)\tablenotemark{b}      & $2.5\times10^{35}$    \\
BB Area (cm$^2$)				       & $8.0\times10^{11}$    \\
$\chi^2_{\nu}$(dof)                   		       & 1.10 (250) \\
\enddata
\tablenotetext{a}{Absorbed flux in units of erg cm$^{-2}$ s$^{-1}$, averaged FPMA \& FPMB values.}
%\tablenotetext{b}{Absorbed hard power-law flux in the 2--30 keV band, averaged FPMA \& FPMB values.}
\tablenotetext{b}{Luminosity is computed for a distance of $3.5$~kpc.}
\tablecomments{Joint fits to FPMA \& FPMB spectra with independent model normalizations. 
Uncertainties are 90\% confidence for three interesting parameters.} 
\end{deluxetable}

The reconstructed \nustar\ coordinates are nominally accurate to
$7\farcs5$ (90\% confidence level), however, a full aspect
reconstruction is not possible for this observation due to the
proximity of the Sun to the target.  For this
pointing, star tracker \#4, which is co-aligned with the X-ray optics,
was not available to recover the absolute aspect or to remove the
$2^{\prime}$ image blur due to telescope mast motion. During the
orbit, a cyclic combination of the other three star trackers was used to
determine the attitude. As outlined below, we analyze data from each
star tracker combination separately to help compensate for the mast motion.

This work uses the data made available via the \nustar\ ToO web page,
which was processed and analyzed using {\tt FTOOLS} 22Oct2018\_V6.25
({\tt NUSTARDAS} 06Jul17\_V1.8.0) with \nustar\ Calibration Database
(CALDB) files of 2016 July~6. Event files were generated for each star
tracker configuration using the {\tt split\_sc} option in {\tt
  nupipeline}. The processed data was uncontaminated by solar wind
events and provides a total of 36~ks of good exposure time.

\subsection{The \nustar\ Analysis}

To generate a \nustar\ spectrum for \xte, we analyzed data separately
from the five available star tracker configurations for this observation,
to allow for a noticeable $\sim$$2^{\prime}$ shift in the source
location on the focal plane, periodic on an orbit timescale. However,
these star tracker configurations break each orbit into five pieces,
adjacent in time, and the drift is not noticeable over these
relatively short intervals.  Thus, we computed the source centroid
on a per orbit, per star tracker configuration basis, to center the
extraction region based on the drift.
%This effectively compensates for the mast induced positional blur of 
%the PSF on the focal plane. 
This allows us to use a smaller extraction aperture of
$0\farcm75$ radius as compared to the $2^{\prime}$ radius of our
previous analysis \citep{got18}, greatly reducing the relative
background contribution in the source aperture at higher energies.

Although the \nustar\ image of \xte\ contained no other sources, there
are regions on the focal plane contaminated by stray light. In
particular, stray light from nearby GX~9$+$1 partially overlaps the
source region in the FPMA image.  For each detector we determined a
suitable background region and accounted for the contaminating
fraction of stray light in the source region, where appropriate.  In
all, for each FPM, we generated 67 sets of source and background
spectral files along with their response functions. The final
summed/averaged FMPA and FPMB spectra and their response files were
grouped to include at least 100 counts per channel and fitted to
several spectral models of interest, as described below.
For these fits, the absorption column is not well-constrained in the
\nustar\ energy band and is held fixed to a nominal $N_{\rm H} =
1.0\times 10^{22}$~cm$^{-2}$. 

\subsubsection{The $<10$~keV \nustar\ Spectrum }

To compare to the early results for the 2003 outburst of \xte,
obtained on 2003 September~8 using \xmm\ \citep{hal05a}, we fit the
\nustar\ spectrum in the restricted 3$-$10~keV band. A blackbody plus
power-law model yields a best-fit temperature $kT = 0.72\pm0.02$~keV
and photon index $\Gamma = 4.2\pm0.2$, with $\chi^2_{\nu} = 1.03$ for
234 degrees-of-freedom (DoF), comparable, but hotter and harder than
that found for the 2003 outburst ($kT = 0.67\pm0.02$~keV, $\Gamma =
3.8\pm0.2$), consistent with the present detection being at an earlier
phase.  The absorbed flux of $F({\rm 2-10~keV}) = (2.13\pm0.06)\times10^{-10}$
erg~s$^{-1}$~cm$^{-2}$, however, is a factor of $\approx$2 greater
than the projected maximum flux of $F({\rm 2-10~keV}) = (0.8-1.1)\times10^{-10}$
erg~s$^{-1}$cm$^{-2}$ for the 2003 outburst \citep{got07}. For the
estimated distance of 3.5~kpc to \xte, the higher bolometric
luminosity of $L_{\rm bol} = 2.6\times 10^{35}$~erg~s$^{-1}$ implies a
blackbody area of $A_{BB} = 9.6\times10^{11}$~cm$^2$, substantially
larger than that measured $\sim$8~months into the previous outburst.

\begin{figure*}[t]
%\vspace{-0.3cm} 
\centerline{
\psfig{figure=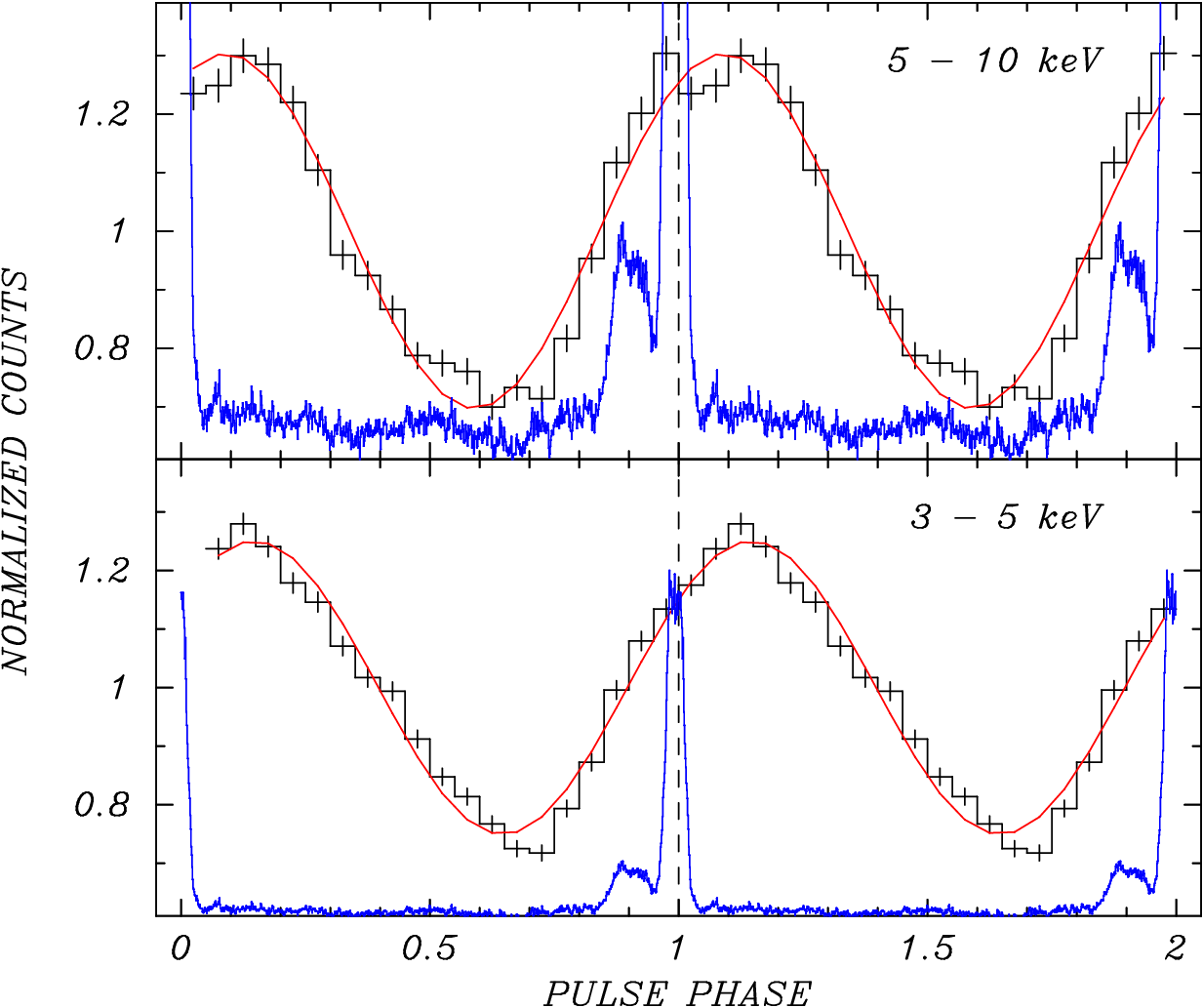,width=0.39\linewidth,angle=270}
\hfill
\psfig{figure=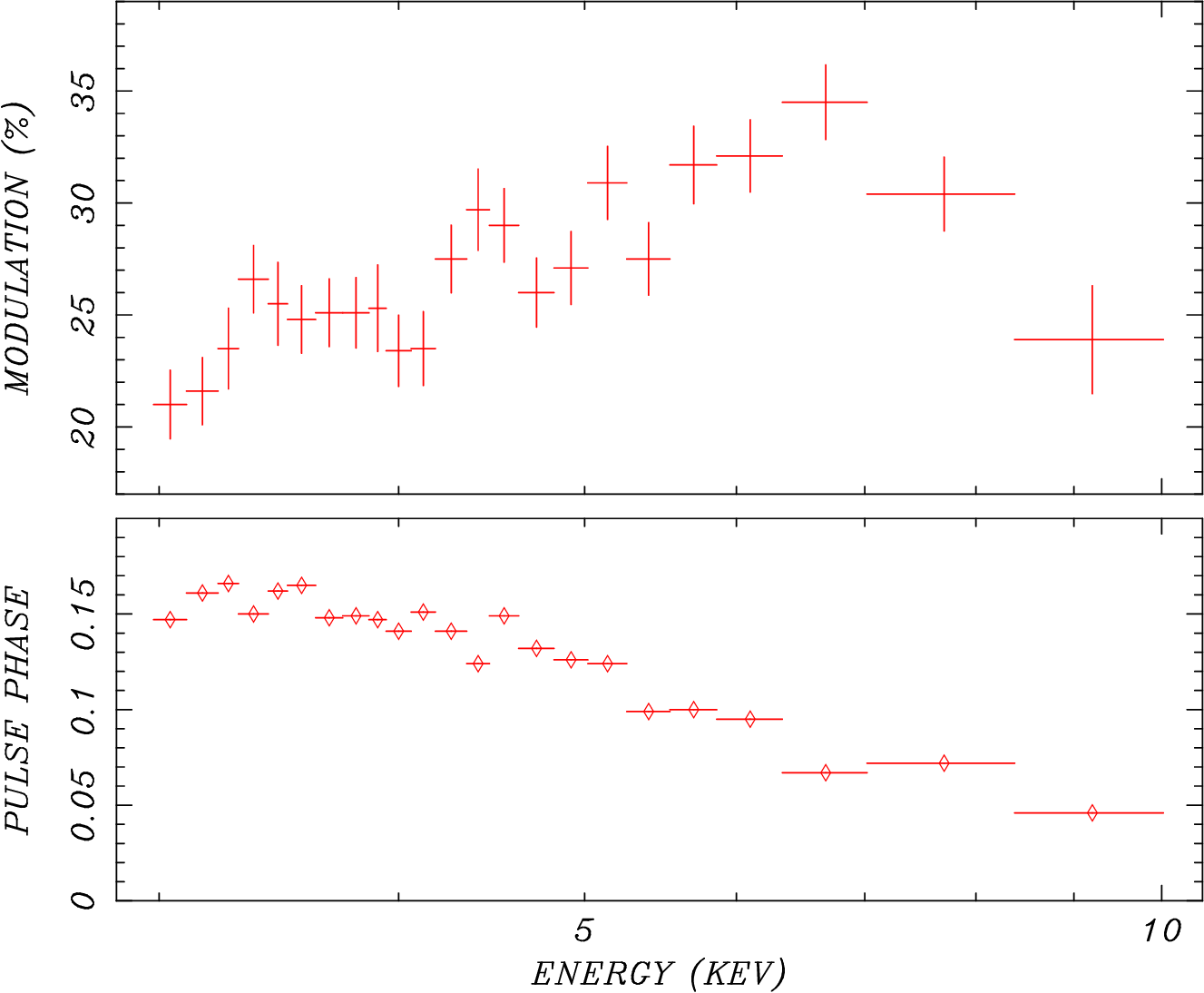,width=0.39\linewidth,angle=270}
\hfill
}
%\vspace{-0.2cm} 
\caption{\footnotesize Left: \nustar\ 3$-$10~keV background
  subtracted pulse profiles of \xte\ (black) in two energy bands folded
  on the radio ephemeris of \citet{low19}, with the best fit sinusoid
  model overlaid (red).  Two cycles are shown for clarity.
  The X-ray observation started at MJD 58465.14 (2018 December 13).
  The blue trace is the radio pulse profile from the December~15
  observation of \citet{dai19} using the Parkes telescope UWL receiver,
  with a center frequency of 2368~MHz and bandwidth of 3328~MHz.
  Phase zero is defined as MJD 58467.943329663 (TDB).
  Two different scalings are shown to display the full dynamic range
  of the radio pulse.
  Right: X-ray pulse modulation (top panel) and phase (bottom panel)
  as a function of energy in uniform bins of $\approx5000$ counts per fold.
%  The dashed line indicates the count-weighted average 3$-$10~keV modulation
%  (top) and phase (bottom) as defined above.
  The slippage in phase with energy accounts for asymmetry
  in the broad-band X-ray pulse profile.}
\label{fig:fold} 
%\vspace{-0.3cm} 
\end{figure*}

In contrast, a perhaps more realistic two-blackbody model, also fitted
in the 3$-$10 keV band, yields much higher temperatures of $kT_{\rm
  hot} = 1.19\pm0.08$~keV and $kT_{\rm warm} = 0.62\pm0.02$~keV
($\chi^2_{\nu} = 1.03$ for 234~Dof) as compared to the prior outburst
($kT = 0.68\pm0.02$~keV and $kT = 0.26\pm0.02$~keV). The absorbed flux
is $F({\rm 2-10~keV}) = (1.9\pm0.3)\times10^{-10}$ for this model. The
bolometric luminosity are $L_{\rm hot} = 6.5\times
10^{35}$~erg~s$^{-1}$ and $L_{\rm warm} = 4.6\times
10^{35}$~erg~s$^{-1}$, with implied areas of $A_{\rm hot} =
3.2\times10^{10}$~cm$^2$ and $A_{\rm warm} = 2.9\times10^{12}$~cm$^2$.
Another physically motivated model is the Comptonized blackbody, with
best-fit parameters $kT = 0.632\pm0.008$~keV and $\alpha = 2.71$
($\chi^2_{\nu} = 1.12$ for 238~Dof). Specifically, we use the model
described in \cite{hal08}, where $\alpha \equiv
-\ln(\tau_{es})/\ln(A)$ is the log ratio of the scattering optical
depth $\tau_{es}$ over the mean amplification $A$ of photon energy per
scattering, valid for $\tau_{es}<<1$ \citep{ryb86}.

\subsubsection{The 3$-$30~keV \nustar\ Spectrum }

The low background contamination of the \nustar\ data set allows us to
model the spectrum up to 30~keV before running out of source photons.
None of the two-component models alone can account for significant
emission evident above 10~keV.  To characterize this
non-thermal emission we fit an additional power-law component to both
the blackbody plus power-law model and the two-blackbody model
described above. As shown in Figure~\ref{fig:spectra}, this yields an
excellent fit for both models.  The resulting best-fit 
parameters, presented in Table~\ref{tab:spectra}, are comparable to
those obtained with the 3$-$10 fits, unchanged within their mutual
90\% confidence levels.  We note that the two-blackbody model yield a
photon index of $\Gamma_{h}=1.5\pm0.5$ for the added power-law
component, a value typical of the non-transient magnetars.  For the
blackbody plus power-law model, however, the index in not well
constrained, and yields an index of $\Gamma_{h}=-0.2\pm1.5$, much
flatter then might be expected \citep[cf.  ][]{eno17}.

We are confident in the need for an additional spectral component to
characterize the $>$10~keV emission, as both FPM detectors yield
consistent results when fitted independently, despite the greater
contamination in the FPMA spectrum due to stray light from GX9+1, as
evident in Figure~\ref{fig:specback}.

\subsubsection{Pulsar Timing \label{timing}}

To search for the expected pulsar signal from \xte, we extracted and
merged barycentered photons from both FPMs using a $0\farcm75$ radius
aperture centered on the source, again compensating for the telescope
mast motion as described above.  Using the $Z^2_1$ statistic, we
recover a highly significant signal with $P=5.5414479(34)$~s at
MJD~58465.14 in the 3$-$10~keV energy range, leaving no doubt as to
the identity of the \nustar\ source, despite its poor aspect.  The
resulting pulse profile is nearly sinusoidal in shape, similar to that
recorded by \xmm\ during the previous outburst. However, the pulse
modulation, determined by fitting a sinusoidal model to the background
subtracted pulse profile, is lower by half (e.g. 21\% vs. 46\% at
3~keV).  The modulation is defined here as the ratio of the sine
amplitude to the average counts per fold bin.

The pulse profile in the 3$-$5~keV and 5$-$10~keV bands is shown in
Figure~\ref{fig:fold} (left) to highlight the clear shift in phase
($\Delta\phi\approx 0.1$) between the two bands, accounting for at
least some of the asymmetry in the broad-band pulse profile.
This is most evident in Figure~\ref{fig:fold} (right), which displays
the phase as a function of energy, suggesting that the emission
components are not strictly co-axial. This figure also shows that the
pulse modulation increases linearly with energy, turning over above
$\sim$7~keV, notably near the spectral component cross-over energy in
Figure~\ref{fig:spectra}.  Within the short \nustar\ exposure, we find
no evidence for bursts on second timescales, as were detected by
\rxte\ during the earlier outburst \citep{woo05}.

Radio pulse observations bracketing the epoch of the \nustar\
observation were obtained by \citet{dai19} using the Parkes Telescope
with the Ultra-Wideband Low receiver system (UWL, \citealt{hob19}),
covering a frequency range of 704--4032~MHz, and by \citet{low19}
using the upgraded Molonglo Observatory Synthesis Telescope
\citep[UTMOST,][]{bai17}.  These were were used to develop an
ephemeris that will be reported in \citet{low19}.  In
Figure~\ref{fig:fold} (left), the X-ray photons were folded on the
radio ephemeris.  A radio pulse profile from \citet{dai19} that was
obtained 2 days after the X-rays is shown in its absolute phase
relation to the X-rays.  The peak of the X-rays lags the radio pulse
by $\approx0.13$ cycles.

Beginning in 2009, after pulsed radio emission turned off, the
spin-down rate of \xte\ was very stable through 2018
\citep{cam16,pin19}. This enables us to extrapolate the \citet{pin19}
ephemeris to the epoch of the current observations.  The predicted
period at the \nustar\ epoch is 5.54146404(74), significantly larger
than the \nustar\ measured value.  This suggests that a large glitch
occurred in conjunction with the (unobserved) onset of the outburst in
late November.  The radio ephemeris give a more precise period than
\nustar, and it suggests a glitch magnitude of $\Delta \nu/\nu =
(4.52\pm0.15)\times10^{-6}$, which is typical of large glitches
observed in AXPs \citep{dib14}.  This is probably a lower limit on the
instantaneous glitch magnitude as it does not take into account any
recovery due to an increase in period derivative between the epoch of
the glitch and that of the first radio observation.

\section{Discussion}

 During the long decay from its 2003 outburst, \xte\ was monitored
closely, revealing an evolving X-ray spectrum \citep{got07}, its
turn-off as a radio pulsar in 2008 \citep{cam16}, its transition to
X-ray quiescence in 2009 \citep{alf16,pin16}, and finally, steady
spin-down until 2018 \citep{pin19}. The first detailed X-ray results
for the 2018 outburst, reported herein, show similarities to this
original event in terms of X-ray flux, spectrum, and pulse properties.
However, the \nustar\ observation was obtained at an earlier phase in
the outburst, within 2-3 weeks of the onset, compared to the 2003
event, when the first comparably useful \xmm\ observations came 8--10
months after onset.

From spectral fits to \xte\ using the blackbody plus power-law model,
the higher bolometric luminosity measured within weeks of the current
outburst results in a larger blackbody area compared to those of the
2003 outburst. Both increases are consistent with catching the
outburst at an earlier epoch. The blackbody temperature is also
consistent with that reported for the 2003 outburst, when it remained
constant for a year before fading linearly \citep{got07}.  However,
for the two-blackbody model, one of temperatures differ significantly
from those reported for the 2003 outburst. Either the two temperatures
are twice as hot as previously recorded or an additional smaller,
hotter emitting region is dominating the spectrum.  The availability
of \nustar\ broadband X-ray image spectroscopy for the current
outburst reveals a hard spectral component above 10~keV not measurable
during the prior outburst, either due to its disappearance a year into
the event or for lack of sensitivity in the \rxte\ scan data at the time.

The measured spectral index favors the two-blackbody model yielding a
value that is similar to that found for a typical magnetar, although
the index measured using the alternative blackbody plus power law model
is not well constrained. However, two concentric thermal hot spots, as
modeled for the 2003 outburst data \citep{per08}, is more consistent
with the \cite{bel09} j-bundle picture in which the hot footpoint is
surrounded by a warm X-ray emitting area.  Future \nustar\
observations will help determine if the hard component is persistent.

The \nustar\ pulse properties differ from those observed during the
previous outburst.  Although the linear increase in modulation with
energy is also evident at the earlier epoch, with roughly the same
slope, the modulation for the newer data is half of that measured
in 2003.  This might be due to the larger area of the present
emitting region, additional flux from the whole NS surface, or a
different location of the emitting region during this early epoch of
the 2018 outburst.  The energy dependence of the modulation and phase
shift of the pulse profile also suggests that the spectral components
are currently not precisely co-aligned, as inferred during the previous outburst
\citep{got05}. This may be transitory, as no phase shift with energy
is evident following the 2003 outburst.

Unlike for the 2003 outburst, for which no regular radio monitoring
was done until years after the X-ray event \citep{hal05b,cam06},
intense radio pulsations heralded the onset of the current X-ray
outburst.  Given the sensitivity of the sparse VLA observations at the
location of \xte\ compiled by \cite{hal05b}, it is entirely possible
that radio pulsations did commence in concert with the 2003 X-ray
outburst.  The current outburst behavior further supports such a
scenario.  Pulsar timing before and after the outburst also suggests
that a glitch occurred in concert with the reactivation.  This would
not have been known of the 2003 outburst, as the pulsar was previously
undiscovered.  It is yet to be determined if the radio emission will
fade smoothly, as for the 2003 outburst, or turn and off at intervals
during the X-ray decay, as is found for other transient radio
magnetars.

Unlike ordinary pulsars, it is not known if radio emission from
magnetars occurs on relatively stable, open dipole field lines, or on
twisted, closed magnetic field-line j-bundles that conduct large
currents and heat the NS surface to X-ray temperatures.  Also, since
outbursts can be triggered by crustal fractures, it is not clear that
each new outburst must originate at the same location on the star.
The radio phase relationship to the X-ray pulse, and the polarization
swing of the radio pulse, may constrain the geometry and location of
the emission regions as was attempted in \citet{cam07b} and
\citet{kra07}.  More than 3 years after the 2003 outburst, the phase
alignment of the X-ray and radio pulse were compared in
\citet{cam07b}, showing that their peaks almost coincide, with the
X-ray lagging the radio by $\approx0.1$ cycles.  With much better
X-ray statistics early in the current outburst, we have established
that the X-ray pulse lags the radio peak by $\approx0.13$ cycles,
consistent with their behavior in the 2003 outburst.  This suggests
that the macroscopic geometry of the magnetic field associated with
the emission did not change between outbursts.

One way to get a lag of the X-rays is for the radio emission to come
from closed magnetic field lines on which there are currents heating
the surface.  In this case, the beaming of the radio emission along
the curved magnetic field lines could account for the lag.
Alternatively, if the radio emission comes from open field lines, at
heights comparable to the radius of the light cylinder, then an X-ray
lag would be expected from both light travel time effects and
relativistic aberration.  In the current outburst it is not yet
possible to model the magnetic geometry and viewing angle using the
rotating vector model of radio polarization.  The early radio emission
is highly erratic in degree and direction of polarization
\citep{dai19}.  If the radio flux persists, its polarization may
settle down to allow the locations of the radio and X-ray emitting
regions to be examined and compared to the results of the 2003
outburst.  The role of open field lines versus closed field lines as
the location of radio emission in magnetars, can be investigated well
using \xte, and is an important subject of exploration during the
current outburst.

\acknowledgements The \nustar\ mission is a project led by the
California Institute of Technology, managed by the Jet Propulsion
Laboratory, and funded by the National Aeronautics and Space
Administration.  We thank the \nustar\ Director for timely approval of
this observation of \xte, and the \nustar\ Operations, Software, and
Calibration teams for support with the execution and analysis of the
observation.  We thank the UTMOST team and the University of Sydney
for maintaining the Molonglo Observatory. E.V.G.  acknowledges NASA
ADAP Grant NNX16AF30G. M.B. acknowledges ARC grant LF150100148.  This
research made use of the \nustar\ Data Analysis Software (NuSTARDAS)
jointly developed by the ASI Science Data Center (ASDC, Italy) and the
California Institute of Technology (USA).  This research also made use
of data and software provided by the High Energy Astrophysics Science
Archive Research Center (HEASARC), which is a service of the
Astrophysics Science Division at NASA/GSFC and the High Energy
Astrophysics Division of the Smithsonian Astrophysical Observatory.
We also acknowledge use of the Astronomer's Telegram (ATel) and the
NASA Astrophysics Data Service (ADS).

\medskip 

%Facilities: \textit{NuSTAR, MAXI}

%\bibliographystyle{}

%\renewcommand{\baselinestretch}{1.0}


\begin{thebibliography}{}

\bibitem[Albano et al.(2010)]{alb10} Albano, A., Turolla, R.,  Israel, G.~L., et al.\ 2010, ApJ, 722, 788 

\bibitem[Alford \& Halpern(2016)]{alf16} Alford, J. A. J., \& Halpern, J. P. 2016, ApJ, 818, 122

%\bibitem[An et al.(2012)]{an12} 
%An, H., Kaspi, V.~M.,  Tomsick, J.~A., et al.\ 2012, ApJ, 757, 68 

%\bibitem[An et al.(2013)]{an13} 
%An, H., Kaspi, V.~M.,  Archibald, R., \& Cumming, A.\ 2013, ApJ, 763, 82

%\bibitem[Anderson et al.(2012)]{and12} 
%Anderson, G.~E.,  Gaensler, B.~M., Slane, P.~O., et al.\ 2012, ApJ, 751, 53 

\bibitem[Arnaud(1996)]{arn96} Arnaud, K.~A.\ 1996, in ASP Conf. Ser. 101, Astronomical Data Analysis Software and Systems V, ed. G. Jacoby \& J. Barnes (San Francisco: ASP), 17

\bibitem[Bailes et al.(2017)]{bai17} Bailes, M., Jameson, A., Flynn, C. et al.\ 2017, PASA, 34, 45

\bibitem[Beloborodov(2009)]{bel09} Beloborodov A.~M.\ 2009, ApJ, 703, 1044 

\bibitem[Beloborodov(2013)]{bel13} Beloborodov A.~M.\ 2013, ApJ, 762, 13 

\bibitem[Bernardini et al.(2009)]{ber09} Bernardini, F., Israel, G.~L., Dall'Osso, S., et al.\ 2009, AAP, 498, 195 

\bibitem[Bernardini et al.(2011)]{ber11} Bernardini, F., Perna, R., Gotthelf, E. V., et al.\ 2011, MNRAS, 418, 638

%\bibitem[Camero et al.(2014)]{cam14} 
%Camero, A., Papitto, A., Rea, N., et al.\ 2014, MNRAS, 438, 3291 

\bibitem[Camilo et al.(2006)]{cam06} Camilo, F., Ransom, S.~M., Halpern, J.~P., et al.\ 2006, Natur, 442, 892 

\bibitem[Camilo et al.(2007a)]{cam07a} Camilo, F., Ransom, S.~M., Halpern, J.~P., Reynolds, J. 2007a, ApJ, 666, L93

\bibitem[Camilo et al.(2007b)]{cam07b} Camilo, F., Cognard, I., Ransom, S.~M., et~al.\ 2007b, ApJ, 663, 497

%\bibitem[Camilo et al.(2015)]{cam15} 
%Camilo, F., Ransom, S.~M., Halpern, J.~P., et al.\ 2015, ApJ, submitted

\bibitem[Camilo et al.(2016)]{cam16} Camilo, F., Ransom, S.~M., Halpern, J.~P., et al.\ 2016, ApJ, 820, 110 

%\bibitem[Coti Zelati et al.(2015)]{cot15} 
%Coti Zelati, F., Rea, N., Papitto, A., et al.\ 2015, MNRAS, 449, 2685 

\bibitem[Coti Zelati et al.(2018)]{cot18} Coti Zelati, F., Rea, N. Pons, J.~A., Campana, S., Esposito, P. 2018, MNRAS, 474, 961	

\bibitem[Dai et al.(2019)]{dai19} Dai, S., Lower, M. E., Bailes, M., et al. 2019, ApJ, submitted, arXiv:1902.04689

%\bibitem[den Hartog \etal(2004)]{den04} den Hartog, P. R., \etal\ 2004, ATel, 293, 1 

%\bibitem[Demorest et al.(2010)]{dem10} 
%Demorest, P.~B., Pennucci, T., Ransom, S.~M., Roberts, M.~S.~E., \& Hessels, J.~W.~T.\ 2010, Nature, 467, 1081 

\bibitem[Desvignes et al.(2018)]{des18} Desvignes, G., Eatough, R., Kramer, M., et al. 2018, ATel, 12285

\bibitem[Dib \& Kaspi(2014)]{dib14} Dib, R., \& Kaspi, V.~M.\ 2014, ApJ, 7484, 37

%\bibitem[Dib et al.(2012)]{dib12} 
%Dib, R., Kaspi, V.~M., Scholz, P., \& Gavriil, F.~P.\ 2012, ApJ, 748, 3 

\bibitem[Durant \& van Kerkwijk(2006)]{dur06} Durant, M., \& van Kerkwijk, M.~H.\ 2006, \apj, 650, 1070 

%\bibitem[Esposito et al.(2009)]{esp09} 
%Esposito, P., Tiengo, A., Mereghetti, S., et al.\ 2009, ApJL, 690, L105 

\bibitem[Eatough et al(2013)]{eat13} Eatough, R.~P., Falcke, H., Karuppusamy, R.  et al. 2013, Nature, 501, 391

\bibitem[Enoto et al(2017)]{eno17} Enoto~T., Shibata, S., Kitaguchi, T., et al. 2017, ApJS, 231, 8 

%\bibitem[Esposito et al.(2011)]{esp11} 
%Esposito, P., Israel, G.~L., Turolla, R., et al.\ 2011, MNRAS, 416, 205 

%\bibitem[Esposito et al.(2013)]{esp13} 
%Esposito, P., Tiengo, A., Rea, N., et al.\ 2013, MNRAS, 429, 3123 

\bibitem[Esposito, Rea \& Israel(2018)]{esp18} Esposito, P., Rea, N. \& Israel, G.~L. \ 2018, arXiv:180305716

\bibitem[Gavriil et al(2002)]{gav02} Gavriil, F.~P., Kaspi, V.~M.; Woods, P.~M.\ 2002, Nature, 419, 142

\bibitem[Gotthelf et al.(2004)]{got04} Gotthelf, E.~V., Halpern, J.~P., Buxton, M., Bailyn, C.  2004, ApJ, 605, 368

\bibitem[Gotthelf \& Halpern(2005)]{got05} Gotthelf, E.~V., \& Halpern, J.~P.\ 2005, ApJ, 632, 1075 

\bibitem[Gotthelf \& Halpern(2007)]{got07} Gotthelf, E.~V., \& Halpern, J.~P.\ 2007, APSS, 308, 79 

\bibitem[Gotthelf et al.(2018)]{got18} Gotthelf, E.~V., Halpern, J. P., Grefenstette, B. W. et al. \ 2018, ATel, 12297

%\bibitem[G{\"u}ver et al.(2007)]{guv07} G{\"u}ver, T., {\"O}zel, F., G{\"o}{\v g}{\"u}{\c s}, E., \& Kouveliotou, C.\ 2007, ApJL, 667, L73 

\bibitem[G{\"u}ver  et al.(2019)]{guv19} G{\"u}ver, T. Majid, W., Enoto, T., et al. 2019, ATel, 12297

\bibitem[Halpern \& Gotthelf(2005)]{hal05a} Halpern, J.~P. \&  Gotthelf, E.~V.\ 2005, ApJ, 618, 874 

\bibitem[Halpern et al.(2005)]{hal05b} Halpern, J.~P., Gotthelf, E.~V., Becker, R.~H., Helfand, D.~J., \& White, R.~L.\ 2005, ApJL, 632, L29 

\bibitem[Halpern \etal(2008)]{hal08} Halpern, J.~P., Gotthelf, E.~V., Reynolds, J., Ransom, S.~M., Camilo, F. 2008, ApJ, 676, 1178 

\bibitem[Harrison et al.(2013)]{Harrison2013} Harrison, F.~A., Craig, W.~W., Christensen, F.~E. et al.~2013, \apj, 770, 103   

\bibitem[Hobbs et al.(2019)]{hob19}
Hobbs, G., et al.\ 2019, in preparation

\bibitem[Hori et al.(2018)]{hor18} Hori, T., Shidatsu, M., Ueda, Y., \etal\ 2018, \apjs, 235, 7

\bibitem[Ibrahim et al.(2004)]{ibr04} Ibrahim, A.~I., Markwardt, C.~B., Swank, J.~H., et al.\ 2004, ApJL, 609, L21

%\bibitem[Kargaltsev et al.(2012)]{kar12} 
%Kargaltsev, O., Kouveliotou, C., Pavlov, G.~G., et al.\ 2012, ApJ, 748, 26

\bibitem[Kaspi et al.(2003)]{kas03} Kaspi, V. M., Gavrill, F. P., Woods, P. M., et al. 2003, ApJ, 588, L93

%\bibitem[Kaspi et al.(2014)]{kas14} 
%Kaspi, V.~M., Archibald, R.~F., Bhalerao, V., et al.\ 2014, ApJ, 786, 84

\bibitem[Kaspi \& Beloborodov(2017)]{kas17} Kaspi, V.~M. \& Beloborodov, A.~.M 2017, ARA\&A, 55, 261

\bibitem[Kramer et al.(2007)]{kra07} Kramer,M., Stappers, B. W., Jessner, A., Lyne, A. G., \& Jordan, C. A.
2007, MNRAS, 377, 107

%\bibitem[Kuiper et al.(2012)]{kui12} 
%Kuiper, L., Hermsen, W., den Hartog, P.~R., \& Urama, J.~O.\ 2012, ApJ, 748, 133 

\bibitem[Levin et al.(2010)]{lev10} Levin, L., Bailes, M., Bates, S., et al.\ 2010, ApJL, 721, L33 

\bibitem[Lower et al.(2018)]{low18} Lower, M., Bailes, M., Jameson, A, et al. 2018, ATel, 12288

\bibitem[Lower et al.(2019)]{low19} Lower, M., et al. 2019, in preparation. 

\bibitem[Lyne et al.(2018)]{lyn18} Lyne, A., Levin, L., Stappers, B., et al. 2018, ATel, 12284

\bibitem[Madsen et al.(2015)]{Madsen15} Madsen, K.~K., Harrison, F.~A., Markwardt, C.~B., et al.~2015, ApJS, 220, 8

\bibitem[Matsuoka et al.(2009)]{mat09} Matsuoka, M., Kawasaki, K., Ueno, S., et al. 2009, PASJ, 61, 999

%\bibitem[Mereghetti et al.(2015)]{mer15} 
%Mereghetti, S., Pons, J.~A., \& Melatos, A.\ 2015, SSR, 26

\bibitem[Mihara et al.(2011)]{mih11} Mihara, T., Negoro, H., Kawai, N., et al. 2011, PASJ. 63S. 623

\bibitem[Mihara et al.(2018)]{mih18} Mihara, T., et al. 2018, ATel, 12291

\bibitem[Minter et al.(2008)]{min08} 
Minter, A.~H., Camilo, F., Ransom, S.~M., Halpern, J.~P., \& Zimmerman, N.\ 2008, ApJ, 676, 1189 

%\bibitem[Molkov \etal(2004)]{mol04} Molkov, S. V., \etal\ 2004, Astron. Lett., 30, 534 

\bibitem[Mori et al.(2014)]{Mori14} Mori, K., Gotthelf, E.~V., Dufour, F., et al.~2014, \apj, 793, 88

%\bibitem[Morrison \& McCammon(1983)]{mor83} 
%Morrison, R., \& McCammon, D.\ 1983, ApJ, 270, 119

\bibitem[Negoro et al.(2016)]{neg16} Negoro, H., Kohama, M., Serino,  M., Saito, H., Takahashi, T., Miyoshi, S, et al. 2016, PASJ, 68, S1


%\bibitem[Olausen \& Kaspi(2014)]{ola14} 
%Olausen, S.~A., \& Kaspi, V.~M.\ 2014, ApJs, 212, 6 

\bibitem[Perna \& Gotthelf(2008)]{per08} Perna, R. \& Gotthelf, E. V. 2008, ApJ, 681, 522

\bibitem[Pintore et al.(2016)]{pin16} Pintore, F., Bernardini, F., Mereghetti, S., et al. 2016, MNRAS, 458, 2088

\bibitem[Pintore et al.(2019)]{pin19} Pintore, F., Mereghetti, A., Esposito, P., et al. 2019, MNRAS, 483, 3832

%\bibitem[Rybicki \& Lightman(1979)]{ryb79} 
%Rybicki, G.~B., \& Lightman, A.~P.\ 1979, Radiative Processes in Astrophysics (New York: Wiley-Interscience)  
\bibitem[Rea \& Esposito(2011)]{rea11} Rea, N., \& Esposito, P. 2011, in ``High-Energy Emission from Pulsars and their Systems'', Astrophysics and Space Science Proceedings, ISBN 978-3-642-17250-2. Springer-Verlag Berlin Heidelberg, 21, 247

%\bibitem[Rea et al.(2012)]{rea12} 
%Rea, N., Pons, J.~A., Torres, D.~F., \& Turolla, R.\ 2012, ApJL, 748, L12 

%\bibitem[Rea et al.(2013)]{rea13} 
%Rea, N., Israel, G.~L., Pons, J.~A., et al.\ 2013, ApJ, 770, 65 

%\bibitem[Rea et al.(2014)]{rea14} 
%Rea, N., Vigan{\`o}, D., Israel, G.~L., Pons, J.~A., \& Torres, D.~F.\ 2014, ApJL, 781, L17 

%\bibitem[Revnivtsev \etal(2004)]{rev04} Revnivtsev, M. G., \etal\ 2004, Astron. Lett. 30, 382 

\bibitem[Rybicki \&  Lightman(1986)]{ryb86} Rybicki, G. B. \& Lightman, A. P. 1986, ``Radiative Processes in Astrophysics, '', Wiley-VCH ISBN 0-471-82759-2. 

\bibitem[Shannon \& Johnston(2013)]{sha13} Shannon, R.~M., \& Johnston, S. 2013, MNRAS, 435, L29

%\bibitem[Scholz et al.(2014)]{sch14} 
%Scholz, P., Kaspi, V.~M., \& Cumming, A.\ 2014, ApJ, 786, 62 

%\bibitem[Szary et al.(2015)]{sza15} 
%Szary, A., Melikidze, G.~I., \& Gil, J.\ 2015, ApJ, 800, 76 

%\bibitem[Tiengo et al.(2013)]{tie13} Tiengo, A.,  et al. 2013, Nature, 500, 312

%\bibitem[Tam et al.(2006)]{tam06} 
%Tam, C.~R., Kaspi, V.~M., Gaensler, B.~M., \& Gotthelf, E.~V.\ 2006, ApJ, 652, 548 

%\bibitem[Torii et al.(1998)]{tor98} 
%Torii, K., Kinugasa, K., Katayama, K., Tsunemi, H., \& Yamauchi, S.\ 1998, ApJ, 503, 843 

%\bibitem[Verner et al.(1996)]{ver96} Verner, D.~A., Ferland, G.~J., Korista, K.~T., Yakovlev, D.~G.~1996, \apj, 465, 487

%\bibitem[Vigan\`o et al.(2013)]{vig13}
%Vigan\`o, D., Rea, N., \&  Pons, J.~A. 2013, MNRAS, 434, 123

%\bibitem[Vurgun et al.(2019)]{vur19} Vurgun, E., et al.  2019, NewA, 67, 45

%\bibitem[Wilms et al.(2000)]{wil00} Wilms, J., Allen, A., \& McCray, R.~2000, \apj, 542, 914

\bibitem[Woods et al.(2005)]{woo05} Woods, P.~M., et al. 2005, ApJ, 629, 985
 
%\bibitem[Zhou et al.(2014)]{zho14} Zhou, P., Chen, Y., Li, 
%X.-D., et al.\ 2014, ApJL, 781, L16               

\end{thebibliography}
\end{document}